# WRITHING DYNAMICS OF CABLES WITH SELF-CONTACT

S. Goyal and N.C. Perkins
University of Michigan, Department of Mechanical Engineering
2350 Hayward, Ann Arbor MI-48109-2125 (U.S.A.)
ncp@umich.edu

Christopher L. Lee
Lawrence Livermore National Laboratory, New Technologies Engineering Division
7000 East Ave., Livermore, California 94550 (U.S.A.)
lee72@llnl.gov

*Abstract*

*Marine cables under low tension and torsion on the sea floor can form highly contorted three-dimensional geometries that include loops (e.g. hockles) and tangles. These geometries arise from the conversion of torsional strain energy to bending strain energy or, kinematically, a conversion of twist to writhe. A dynamic form of Kirchhoff rod theory is reviewed herein that captures these nonlinear dynamic processes. The resulting theory is discretized using the generalized-α method for finite differencing in both space and time. Numerical solutions are presented for an example system of a cable subjected to increasing twist at one end. The solutions show the dynamic evolution of the cable from an initially straight element, through a buckled element in the approximate form of a helix, through the dynamic collapse of this helix into a loop, and subsequent intertwining of the loop with multiple sites of self-contact.*

## 1. INTRODUCTION

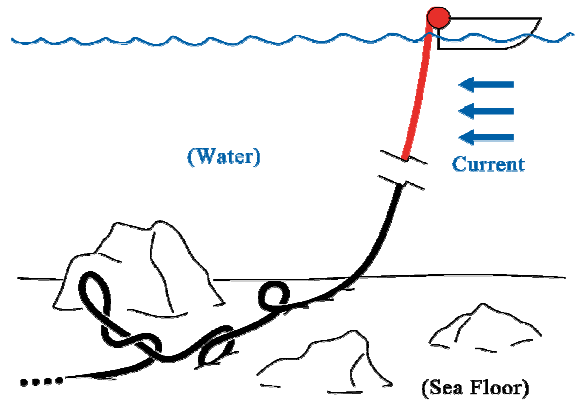

Figure 1: Low tension cable forming loops and tangles on the sea floor.

Cables laid upon the sea floor may form loops and tangles as illustrated in Fig. 1. The loops (sometimes referred to as *hockles*) may cause localized damage and, in the case of fiber optic cables, may also prevent signal transmission. These highly nonlinear deformations are initiated by conditions of low cable tension (or slight compression) and torsion sufficient to induce a torsional "buckling" of the cable. The effect of bending and torsional stiffness becomes pronounced in such "low tension zones" of the cable. Several prior investigations of cable loop formation have employed nonlinear equilibrium (static) rod theories to analyze the equilibrium forms of cables under torsion and low tension; see, for example, Coyne [1], Rosenthal [2,3], Liu [4], Tan and Witz [5]. The stability of these equilibrium forms may be assessed using local stability analyses as in Lu and Perkins [6,7]. The overall buckling process, however, is inherently a *dynamic process,* and this fact has recently been recognized by Gatti-Bono and Perkins [8] who employed a nonlinear dynamic rod theory to simulate loop formation in a cable under compression.

Gobat and Grosenbaugh [9] employ a similar rod theory for underwater cables and they discretize their model with the generalized-α method developed by Chung and Hulbert [10] for integration with respect to time. Gobat and Grosenbaugh [11] also illustrate the substantial advantages of the generalized-α method over other methods for cable dynamics simulation. Goyal et al. [12] extended the cable model of Gatti-Bono and Perkins [13] with the generalized-α method in both space and time and employed it to study loop formation under torsion.

In this paper, we extend the model described in Goyal et al. [12] to include the dynamic self-contact that results in intertwining as shown in Fig. 1. We start with a brief review of the model and the numerical discretization. Next, we propose a model for cable self-contact. Then, we illustrate and discuss an example of a non-uniform cable subject to increasing twist. Finally, we close with a summary of our conclusions.

## 2. PHYSICAL MODEL AND NUMERICAL SCHEME

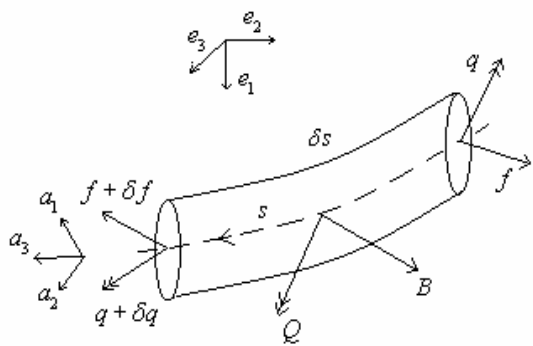

Figure 2: Infinitesimal cable element for formulating equations of motion.

We review here the equations of motions for the cable model and its numerical discretization described in Goyal et al. [12]. All quantities at any spatial point *s* and at time *t* (see Fig. 2), are expressed in the material-fixed, local reference frame $\{a_i\}$ aligned with the centerline tangent and the 'principal torsion-flexure axes' [14]. The dynamic state of any cable cross-section is represented by four vector quantities defined along the cable centerline: the linear and angular velocities of the cross-section, *v* and *ω* respectively, the internal force *f*, and the curvature vector *κ* that generates an internal moment *q* through the assumed linear constitutive law :

$$q = \begin{bmatrix} EJ_1 & 0 & 0 \\ 0 & EJ_2 & 0 \\ 0 & 0 & GJ_3 \end{bmatrix} \kappa \qquad (1)$$

The cable material and geometric parameters in Eq. (1) are defined in Tables 1 and 2. We ignore structural damping and any intrinsic (unstressed) curvature in arriving at Eq. (1).

The four partial differential equations relating the four vector unknowns are :

$$\frac{\partial \omega}{\partial s} + \kappa \times \omega = \frac{\partial \kappa}{\partial t} \tag{2}$$

$$\frac{\partial v}{\partial s} + \kappa \times v = \omega \times a_3 \tag{3}$$

$$\frac{\partial f}{\partial s} + \kappa \times f = \rho_c A_c \left\{ \frac{\partial v}{\partial t} + \omega \times v \right\} - B \tag{4}$$

$$\frac{\partial q}{\partial s} + \kappa \times q = I \frac{\partial \omega}{\partial t} + \omega \times (I\omega) + f \times a_3 - Q \tag{5}$$

Equation (2) is a compatibility condition, Eq. (3) is an inextensibility constraint, and Eqs. (4) and (5) are the Newton-Euler equations of motion for an infinitesimal cable element subject to distributed (body) force $B$ and moment $Q$ by the surrounding environment. $I$ denotes the mass moment of inertia per unit cable length and it is calculated per Table 2.

We discretize the above set of equations in both space and time using the generalized-α method as described in Goyal et al. [12]. This is a 1-step method in both space and time with averaging of coefficients based on a single numerical parameter. The method is unconditionally stable and $2^{nd}$ order accurate in both space and time. The single numerical parameter can be varied to smoothly control maximum numerical dissipation. The difference equations so obtained are implicit and their solution needs to satisfy the boundary conditions at the two ends. They are solved using shooting method in conjunction with Newton-Raphson iteration as described in Goyal et al. [12].

## 3. CABLE SELF-CONTACT

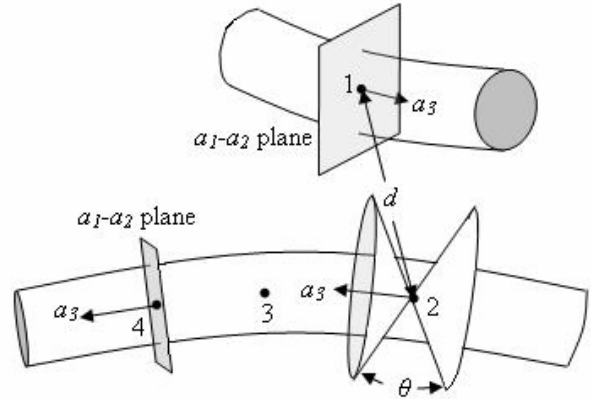

Figure 3: Two cable segments approaching contact.

We will introduce a self-contact model for the cable that is used to simulate loop formation and intertwining. The contact forces are normal to the cable surfaces and allow for sliding contact. Figure 3 shows two segments of the cable that are approaching contact. The upper segment contains one spatial grid point marked as 1, while the lower segment contains three grid points marked as 2, 3 and 4. At each grid point, the $a_1$-$a_2$ plane is orthogonal to the tangent $a_3$. We introduce an *aperture angle* $\theta$ that creates a pair of conical surfaces centered at the grid point as illustrated at point 2 in Fig. 3. Note that this aperture reduces to the $a_1$-$a_2$ plane as $\theta \to 0°$, and it expands to the entire space as $\theta \to 180°$.

We use this aperture to control the number of points that may potentially interact through self-contact. During simulation, the distance $d$ between each pair of grid points on the cable is measured. A repulsive (contact) force is introduced between a pair if and only if two conditions are met: 1) the distance $d$ is within a specified tolerance, and 2) the two grid points lie within each other's aperture. The (distributed) repulsive force is of the form (see Tables 1 and 2 for the cable parameters used below):

$$B_{contact} = \rho_c A_c \left( \frac{k_1}{(d-0.5D)^{k_2}} + \frac{k_3}{d} \frac{\partial d}{\partial t} \left| \frac{\partial d}{\partial t} \right|^{k_4} \right) \tag{6}$$

This force captures both contact stiffness and contact damping. For the example simulation presented below, we have $k_1 = 10^{-7} m^4/s^2$, $k_2 = 3$, $k_3 = 10^{-6}$ and $k_4 = 1$.

## 4. RESULTS AND DISCUSSION

Table 1: Cable and simulation parameters.

| Quantity | Units (SI) | Value |
|---|---|---|
| Young's Modulus, $E$ | Pa | $1.25 \times 10^7$ |
| Shear Modulus, $G$ | Pa | $5.00 \times 10^6$ |
| Cable Diameter, $D$ | m | See Fig. 4 |
| Cable Length, $L$ | m | $1.00 \times 10^0$ |
| Cable Density, $\rho_c$ | Kg/m$^3$ | $1.50 \times 10^3$ |
| Fluid Density, $\rho_w$ | Kg/m$^3$ | $1.00 \times 10^3$ |
| Temporal Step, $\Delta t$ | s | $1.00 \times 10^{-1}$ |
| Spatial Step, $\Delta s$ | m | $1.00 \times 10^{-3}$ |

Table 2: Cable cross-section properties.

| Quantity | Formula | Units (SI) |
|---|---|---|
| Cross-section Area | $A_c = \dfrac{\pi D^2}{4}$ | m$^2$ |
| Area Moments of Inertia (bending) | $J_{1,2} = \dfrac{A_c D^2}{16}$ | m$^4$ |
| Area Moment of Inertia (torsion) | $J_3 = \dfrac{A_c D^2}{8}$ | m$^4$ |
| Mass Moment of Inertia per unit cable length | $I = \rho_c J$ | Kg-m |

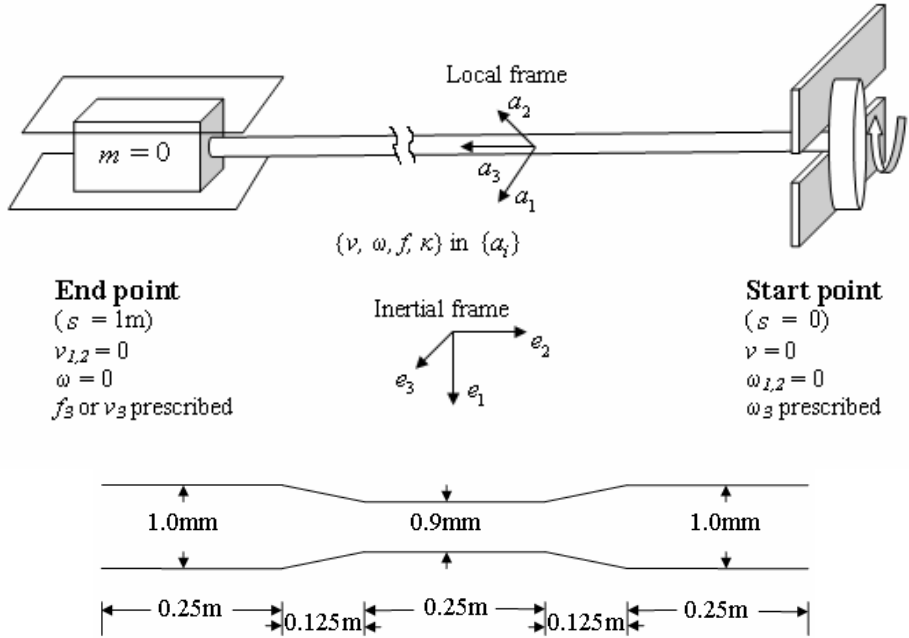

Figure 4: A low tension cable under increasing twist created by rotating the right end. Left end may be free to slide, or have prescribed sliding velocity or reaction (tension).

The model is used to explore several possible dynamic motions that are generated by slowly twisting one end of an elastic cable. The parameters that define the example are listed in Tables 1 and 2 and a schematic of this example is illustrated in Fig. 4. A non-uniform cable is selected with a thinner middle region to localize buckling and loop formation near the center of the cable. The cable diameter $D$ varies over the length as shown in Fig. 4 (not to scale). This small 10% variation in the diameter produces significant ($\approx 35\%$) variation in torsional and bending stiffness.

The integration begins with the cable initially horizontal and stress-free. It is in a fluid (with no flow) that provides added mass and drag (as modeled by a Morison formulation). There is negligible structural damping and no gravity (and hence no buoyancy). However, a minute distributed force in the downward ($e_1$) direction is added to initiate buckling.

The right end (referred to as the "Start point", $s = 0$ in Fig. 4) of the cable is subjected to an increasingly larger rotation about the $a_3$ (tangent) axis. This end cannot move and it is otherwise constrained in rotation (no rotation about the principal axes $a_1$ and $a_2$, i.e. $\omega_1 = \omega_2 = 0$). The left end (referred to as the "End point", $s = L$) of the cable is fully restrained in rotation (about all three axes) and cannot translate in the transverse ($a_1$-$a_2$) plane. This end, however, may translate along the $a_3$ axis. By increasing the rotation at the right end, enough internal torque is induced to generate torsional buckling and subsequent nonlinear dynamic response. This rotation is generated by prescribing the angular velocity component $\omega_3$ at the right end as shown in Fig. 5 (not to scale). The left end is allowed to translate freely during the first 30 seconds and is then held fixed to control what would otherwise be a very rapid collapse.

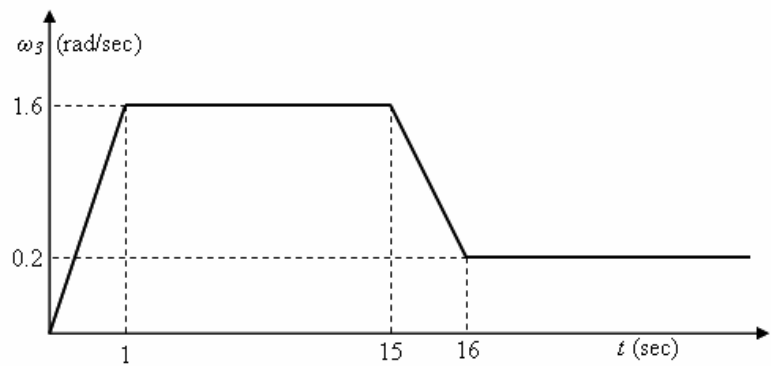

Figure 5: Angular velocity prescribed at the right end.

As the right end is first twisted by a modest amount, the cable remains straight. There is an abrupt change however when the twist reaches a critical value (at approximately 16 seconds) when the Greenhill buckling condition [15] is achieved and the straight (trivial) configuration becomes unstable. The model employed here captures this initial instability as well as the subsequent nonlinear motion that leads to loop formation and intertwining. Figure 6 shows snap-shots of the cable at four different time-steps during the buckling. The geometry just after initial buckling is approximately helical as can be observed in the snap-shot at 20 seconds. Notice that the cable appears to make a single helical turn as expected from the fundamental buckling mode of the (simpler) linearized theory [15].

As the left end is allowed to slide towards the right end, the helical cable undergoes a secondary buckling in which it rapidly collapses in forming a (nearly) planar loop with self-contact. This collapse occurs at approximately 29 seconds in this example. The dynamic collapse is predicted from investigations of the stability of the equilibrium forms of a cable under similar loading conditions; refer to Lu and Perkins [6] and studies cited therein.

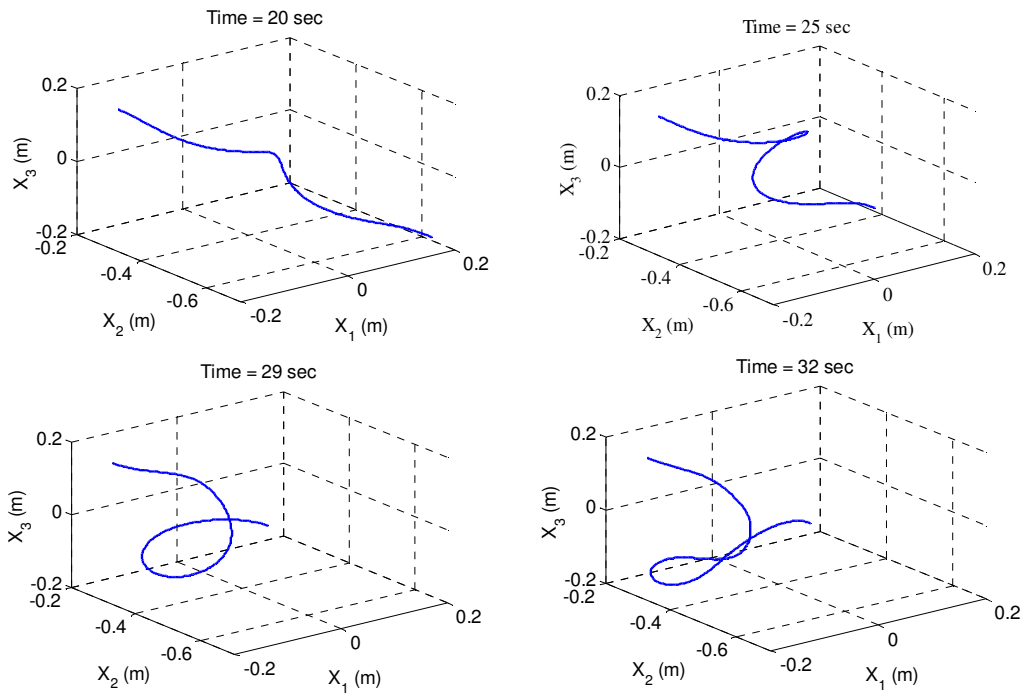

Figure 6: Snap-shots at various time steps during buckling.

The snap-shot at 25 seconds shows the three-dimensional shape of the cable just before dynamic collapse. The center of the cable has rotated approximately 90° about the vertical ($e_1$) axis so that the tangent at this (mid-span) point is now orthogonal to the loading ($e_2$) axis. This was a noted bifurcation condition in Lu and Perkins [6] at which the three-dimensional equilibrium form loses stability. The dynamic collapse thereafter is depicted in the snap-shot at 29 seconds. This nearly planar loop, however, is still unstable and rapidly continues to rotate leading to intertwining with two sites of contact. A snapshot of intertwined cable at 32 seconds is shown.

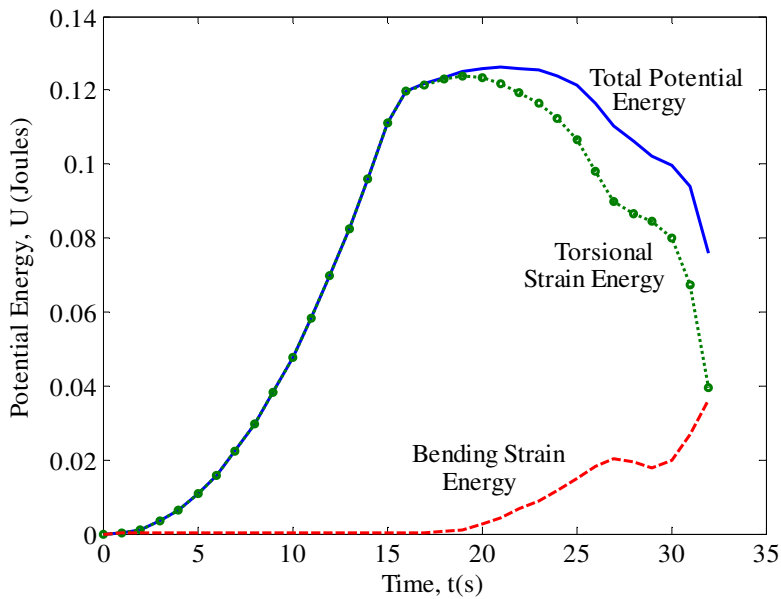

Figure 7: Variation in torsional and bending strain energies during the buckling process.

The entire dynamic collapse depicted in Fig. 6 involves a conversion of torsional strain energy to bending strain energy as shown in Fig. 7. The process begins with an increase in torsional strain energy prior to the collapse from the prescribed rotation at the right end. The maximum torsional strain energy occurs at approximately 18 seconds and follows shortly after the initial buckling (approx. 16 seconds) when bending strain energy first develops. This is followed by a rapid (dynamic) conversion of torsional strain energy to bending strain energy until around 25 seconds when it approaches the second bifurcation depicted in Fig. 6. During secondary bifurcation, the cable loses both torsional and bending strain energies until the first self-contact.

During intertwining, the torsion in the cable is reduced rapidly with a modest increase in curvature. This results in a rapid loss of torsional strain energy and total potential energy with a modest increase in bending strain energy. The loss in total potential energy is accompanied by an increase in kinetic energy suggesting faster dynamics during this stage. This example simulation terminated at 32 seconds due to high velocities.

It is interesting to observe that the topological changes for the cable above are also exhibited by the long chain bio-molecule DNA (Deoxy-ribonucleic acid) during supercoiling. Moreover, supercoiling plays a crucial role in the biological functions of DNA including transcription, replication and compaction inside a cell nucleus or virus. The long-length scale mechanics of DNA have previously been studied using elastic rod theories under equilibrium (static) conditions [16-23]. As noted by Calladine [24], if the longest of human DNA molecule were enlarged to have a width of ordinary kite string, it would extend to about 100km. Given this extreme slenderness ratio and the fact that in-vivo DNA is immersed in an aqueous environment, the analogy to an underwater cable is attractive. The conversion of torsional strain energy to bending strain energy described above, are usually explained *kinematically* for DNA as a conversion of *twist* to *writhe*. We explain this conversion in the above example, starting with definitions for *twist* and *writhe*.

Writhe (Wr) is defined as the number of cross-overs one can see averaged over all possible views of the DNA strand (or cable). For our initially straight configuration (Fig. 4), this quantity is zero, with the first self-contact at 29 seconds (Fig. 6), it is one, and with the subsequent intertwined configuration at 32 seconds, it is two. (Note that if we see a cross-over in three orthogonal views, we will see a cross over in all possible views). The writhe (Wr) is purely a function of the space curve defining the cable centerline and it is positive or negative based on whether it is right-handed or left-handed [24,25].

Twist (Tw) is computed from :

$$Tw = \frac{1}{2\pi} \int_0^L \kappa_3 ds \qquad (7)$$

The sum Tw + Wr equals to the number of rotations of the right boundary in our example and this sum is called the *Linking number* Lk[1].

In our example, the initial twisting phase rapidly introduces Lk $\cong$ 4.0 linking number, all in the form of twist, prior to the initial buckling as shown in Fig. 8. The linking number is then increased slowly thereafter. During buckling, Wr increases to 1.0 when self-contact occurs at 29 seconds and Tw reduces by the same amount so that the sum Wr + Tw = Lk. After the first self-contact, the loop continues to rotate as it intertwines. In doing so, every half rotation of the loop establishes an additional contact site thereby increasing Wr by 1.0 and reducing Tw by 1.0. At 32 seconds, Wr is slightly larger than 2.0. Thus, we observe two crossovers in any three orthogonal views of the snapshot at 32 seconds. There is an equivalent loss in Tw as shown in Fig. 8.

---

[1] This, in general, is not true for boundary conditions that allow rotation about principal axes ($\omega_{1,2} \neq 0$). See Calugareanu [26] and White [27] for proof of conservation of the Linking number (Lk) and refer to Calladine [24] for example discussions for DNA.

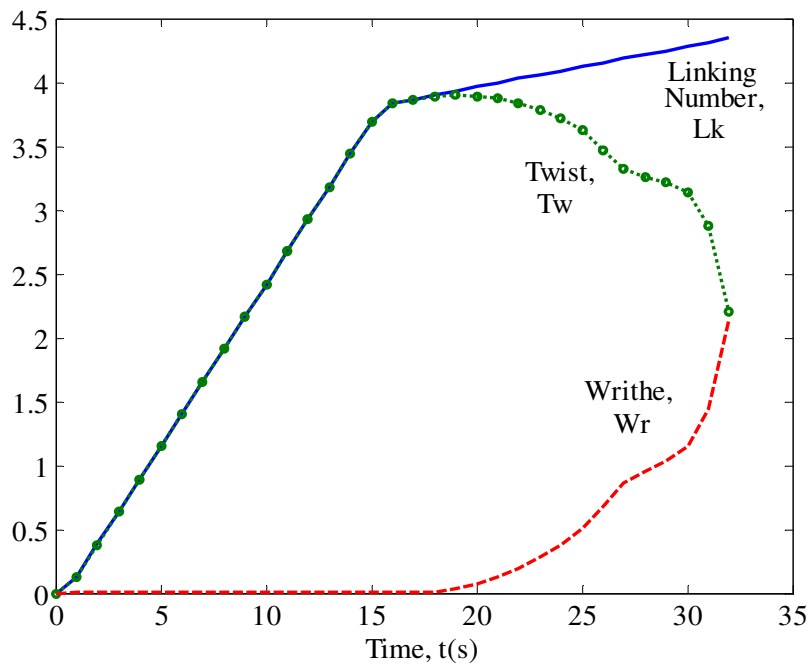

Figure 8: Exchange of twist (Tw) and writhe (Wr). The linking number Lk = Tw + Wr.

It should be noted that bending strain energy is not a measure of writhe (or the number of crossovers), and unlike linking number, total potential energy of the cable is not a conserved quantity. Therefore, the *kinematic* analysis in terms of twist and writhe not only validates the numerical results, but also provides a simple understanding of the topological changes of the cable under torsion.

## 5. SUMMARY AND CONCLUSIONS

This paper reviews a rod theory and a numerical algorithm that can be used to study the nonlinear dynamics of highly contorted cables. While the primary objective is to model the dynamics of marine cables leading to the formation of loops and tangles on the sea floor, it is also recognized that the same techniques may apply to modeling the supercoiled states of DNA and the dynamic transitions between these states. These techniques are used herein to study the response of a prototypical system, composed of an elastic cable subjected to increasing twist. Numerical simulations reveal that the originally straight cable undergoes two bifurcations in succession as twist is added. The first one occurs at the Greenhill buckling condition where the trivial (straight) equilibrium becomes unstable and the cable buckles into the approximate shape of a helix. This helix grows in amplitude with increasing twist. One measure of this growth is the continued rotation of the tangent at the mid-span point. When this tangent becomes orthogonal to the loading axis (axis of the original straight cable), the helix experiences a second bifurcation and collapses dynamically towards a planar loop. The planar loop is unstable and continues to rotate upon itself resulting in intertwining with multiple sites of self-contact. The results also illustrate the possibility of using rod theory to simulate the dynamics of DNA supercoiling over long-length scales. The spatially-varying cable properties can accommodate the sequence-dependent structure of DNA.

ACKNOWLEDGMENTS

The authors gratefully acknowledge the research support provided by the U.S. Office of Naval Research and the Lawrence Livermore National Laboratories. The work of Christopher L. Lee was performed under the auspices of the U.S. Department of Energy by the University of California, Lawrence Livermore National Laboratory under contract No. W-7405-ENG-48.

**REFERENCES**

[1] J. Coyne, "Analysis of the Formation and Elimination of Loops in Twisted Cable", *IEEE Journal of Oceanic Engineering*, Vol. 15, pp. 72-83, 1990.
[2] F. Rosenthal, "The Application of Greenhill's Formula to Cable Hockling", *Journal of Applied Mechanics*, Vol. 43, pp. 681-683, 1976.
[3] F. Rosenthal, "Greenhill's Formula and the Mechanics of Cable Hockling", *NRL Report 7940*, Naval Research Laboratory, Washington D.C., 1975.
[4] F.C. Liu, "Kink Formation and Rotational Response of Single and Multistrand Electromechanical Cables", *Technical Note N-1403*, Civil engineering Laboratory, Naval Construction Battalion Center, Port Hueneme, CA, 1975.
[5] Z. Tan and J.A. Witz, "On the Flexural-Torsional Behavior of a Straight Elastic Bean Subject to Terminal Moments", *Journal of Applied Mechanics*, Vol. 60, pp. 498-505, 1993.
[6] C.L. Lu and N.C. Perkins, "Complex Spatial Equilibria of U-joint Supported Cables under Torque, Thrust and Self-weight", *International Journal of Non-linear Mechanics*, Vol. 30, pp. 271-285, 1995.
[7] C.L. Lu and N.C. Perkins, "Nonlinear Spatial Equilibria and Stability of Cables under Uni-axial Torque and Thrust", *Journal of Applied Mechanics*, Vol. 61, pp. 879-886, 1994.
[8] C. Gatti-Bono and N.C. Perkins, "Dynamic Analysis of Loop Formation in Cables under Compression", *International Journal of Offshore and Polar Engineering*, Vol. 12, pp. 217-222, 2002.
[9] J.I. Gobat and M.A. Grosenbaugh, "Application of the Generalized-Alpha Method to the Time Integration of the Cable Dynamics Equations", *Computational Methods in Applied Mechanics and Engineering*, Vol. 190, pp. 4817-4829, 2001.
[10] J. Chung and G.M. Hulbert, "A time Integration Algorithm for Structural Dynamics With Improved Numerical Dissipation – The Generalized-Alpha Method", *Journal of Applied Mechanics*, Vol. 60, pp. 371-375, 1993.
[11] J.I. Gobat and M.A. Grosenbaugh and M.S. Triantofyllou, "Generalized-Alpha Time Integration Solutions for Hanging Chain Dynamics", *Journal of Engineering Mechanics*, Vol. 128, pp. 677-687, 2002.
[12] S. Goyal, N.C. Perkins and C.L. Lee, "Torsional Buckling and Writhing Dynamics of Elastic Cables and DNA", in *Proceedings of the ASME 2003 Design Engineering and Technical Conference*, to appear.
[13] C. Gatti-Bono and N.C. Perkins, "Physical and Numerical Modeling of the Dynamic Behavior of a Fly Line", *Journal of Sound and Vibration*, Vol. 225, pp. 555-577, 2002.
[14] E.H. Love, *A Treatise on the Mathematical Theory of Elasticity*, Dover Publications, New York (1927), Chapter 18, pp. 281-284.
[15] S.P. Timoshenko and J.M.Gere, *Theory of Elastic Stability*, McGraw-Hill Book Company, New York (1961), Chapter 2, pp. 156-157.
[16] I. Tobias, D. Swigon and B.D. Coleman, "Elastic Stability of DNA Configurations. I. General Theory", *Physical Review E*, Vol. 61, pp. 747-758, 2000.
[17] B.D. Coleman, D. Swigon and I. Tobias, "Elastic Stability of DNA Configurations. II. Supercoiled Plasmids with Self-contact", *Physical Review E*, Vol. 61, pp. 759-770, 2000.

[18] O. Gonzales, J.H. Maddocks, F. Schuricht and H. von der Mosel, "Global Curvature and Self-contact of Nonlinearly Elastic Curves", *Calculus of Variations and Partial Differential Equations*, Vol. 14, pp. 29-68, 2002.
[19] R.S. Manning, J.H. Maddocks and J.D. Kahn, "A Continuum Rod Model of Sequence-dependent DNA Structure", *Journal of Chemical Physics*, Vol. 105, pp. 5626-5646, 1996.
[20] T. Schlick and G. Ramachandranan, "Buckling transitions in Superhelical DNA: Dependence on the Elastic Constants and DNA size", *Biopolymers*, Vol. 41, pp. 5-25, 1997.
[21] T. Schlick, "Modeling Superhelical DNA: Recent Analytical and Dynamic Approaches", *Current Opinion in Structural Biology*, Vol. 5, pp. 245-262, 1995.
[22] W.K. Olson and V.B. Zhurkin, "Modeling DNA deformations", *Current Opinion in Structural Biology*, Vol. 10, pp. 286-297, 2000.
[23] B. Fain, J. Rudnick and S. Ostlund, "Conformations of linear DNA", *Physical Review E*, Vol. 55, pp. 7364-7368, 1997.
[24] C.R. Calladine and H.R. Drew, *Understanding DNA, the molecule and how it works*, Academic Press, New York (1997), chapter 1, pp. 8 and chapter 6, pp. 120-144.
[25] F.B. Fuller, "Writhing Number of a Space Curve", in *Proceedings of the 1971 National Academy of Sciences of the United States of America*, Vol. 68 pp. 815-819.
[26] G. Călugăreanu, "Sur les classes d'isotopie des noeuds tridimensionnels et leurs invariants", *Czechoslovak Math. J.*, Vol. 11, pp. 588-625, 1961.
[27] J.H. White, "Self-Linking and Gauss-Integral in Higher Dimensions", *American Journal of Mathematics*, Vol. 91, pp. 693-728, 1969.